# GRAVITATIONAL LENSING AND ANISOTROPIES OF CBR ON THE SMALL ANGULAR SCALES

J.M. Liu[a]   J.G. Gao[b]

[a] *International School for Advanced Studies(SISSA),*

*Via Beirut n.2-4, 34013 Trieste Italy*

[b] *I.C.R.A., Dipartimento di Fisica, Università "La Sapienza",*

*Piazzale Aldo Moro, 5-00185 Roma Italy*

We investigate the effect of gravitational lensing, produced by linear density perturbations, for anisotropies of the Cosmic Background Radiation (CBR) on scales of arcminutes. In calculations, a flat universe ($\Omega = 1$) and the Harrison-Zel'dovich spectrum ($n = 1$) are assumed. The numerical results show that on scales of a few arcminutes, gravitational lensing produces only negligible anisotropies in the temperature of the CBR. Our conclusion disagrees with that of Cayón *et al.* who argue that the amplification of $\Delta T/T$ on scales $\leq 3'$ may even be larger than 100%.

## I. INTRODUCTION

Many authors have computed the effect of gravitational lensing for anisotropies of the CBR (Blanchard & Schneider 1987; Kashlinsky 1988; Cole & Efstathiou 1989; Tomita & Watanabe 1989; Linder 1990; Watanabe & Tomita 1991; Feng & Liu. 1992; Cayón *et al.* 1993). Unfortunately, however, their conclusions are controversial. Roughly speaking, there exist three different kinds of conclusion so far. The first is that gravitational lensing effects strongly erase fluctuations of the CBR on scales of a few arcminutes (Kashlinsky 1988); the second is that an appreciable, even strong, amplification of $\Delta T/T$ is possible (Sasaki 1989; Linder 1990; Cayón *et al.* 1993); the last is that gravitational lens effects on the CBR are negligible (Cole & Efstathiou 1989; Tomita & Watanabe 1989).

Recently, Cayón *et al.* presented calculations of the gravitational lensing effects, produced by linear density fluctuations on the CBR and got an interesting result. Their work implied that there should be an appreciable amplification, of the order of 20%, for $\Delta T/T$ in present experiments on the scales of several arcminutes, whereas previous work including the effect of nonlinear density fluctuations found a negligible amplification (Cole & Efstathiou 1989). This implies that the gravitational lensing effect on anisotropies of the CBR due to linear density fluctuations overwhelms that due to nonlinear clustering. This result is surprising and difficult to interpret.

In this paper, we use a new formalism to calculate the effect of gravitational lensing on anisotropies of the CBR produced by linear density fluctuations. Our calculations show that the gravitational lensing effect on scales of a few arcminutes is essentially negligible in contrast with the results of Cayón *et al.*

Within the geometrical optics approximation, a gravitational field is equivalent to an optical medium with a refractive index different from unity, and the deflection of light may be interpreted in terms of the refractive index and its spatial variation. The amplitude and phase fluctuations, produced by a random gravitational field, may then be calculated using the usual methods of random medium optics (Fang 1982).

In Sec.II, we establish the basic equations for a wave scattered by a gravitational potential and for the amplitude of the scattered wave. In Sec.III, we present the explicit formulae for calculating the anisotropies of the CBR. Numerical results and brief conclusions are summarized in Sec.IV.

## II. THE BASIC METHOD

Under some reasonable assumptions (Cole & Efstathiou 1989), one can prove that there exists a gauge such that the metric perturbations are characterized by a single potential $\phi(t,\mathbf{x}) \ll 1$

$$ds^2 = -(1+2\phi)dt^2 + (1-2\phi)a^2(t)\delta_{ij}dx^i dx^j, \qquad (1)$$

where $a(t)$ is the scale factor of the universe. The relationship between the gravitational potential $\phi$ and the matter density perturbation $\delta\rho$ is with $8\pi G = c = 1$

$$\Delta\phi = \frac{1}{2}a^2 \rho_b \delta\rho. \qquad (2)$$

In the background universe for the linear perturbation (which we assume to have $\Omega_0 = 1$), $\phi$ is time-independent.

For the sake of simplicity in writing the Maxwell equations in a perturbed metric, we use the conformal time $\tau = 3\sqrt{3}t^{1/3}$ instead of $t$. Then

$$ds^2 = -\frac{1}{3}(1+2\phi)\tau^4 d\tau^2 + \frac{1}{3}(1-2\phi)\tau^4 \delta_{ij}dx^i dx^j. \tag{3}$$

The speed of light in the metric of Eq.(3) is set to unity. A fundamental problem for treating light propagating in an inhomogeneous universe is that of how to describe the gravitational lensing effect. According to geometrical optics, a gravitational field is equivalent to a medium with a refractive index different from unity and the deflection of light may be interpreted in terms of the refractive index and its spatial variation. As a simple version of this analogue, consider the spatial part of the photon four-momentum $k^\mu = \frac{dx^\mu}{d\lambda}$ to be directed along the $x^3$ axis; one then has the following geodesic equation:

$$dk^i = -2\frac{\partial \phi}{\partial x^i}dx^3, \qquad i=1,2. \tag{4}$$

On the other hand, the change in the direction of a light ray propagating in an inhomogeneous medium with refractive index $n = 1 + n'$ is

$$dk^i = \frac{\partial n'}{\partial x^i}dx^3, \qquad i=1,2. \tag{5}$$

By comparing of Eqs.(4) and (5), it is obvious that the effect of the perturbed gravitational field is equivalent to a change in the refractive index, i.e.

$$n' = -2\phi. \tag{6}$$

Therefore, considering the equivalence of the two descriptions of light propagating in an inhomogeneous matter distribution and in a medium with inhomogeneous refractive index, the Maxwell equations may be written in the following form

$$\frac{\partial \mathbf{E}}{\partial \tau} = (1+2\phi) \nabla \times \mathbf{H}, \tag{7}$$

$$\frac{\partial \mathbf{H}}{\partial \tau} = -(1 + 2\phi) \nabla \times \mathbf{E}. \qquad (8)$$

Formally, adopting Hanni's approach (Hanni 1977), the Maxwell equations in an universe with an inhomogeneous matter distribution may be obtained in more rigorous way.

The propagation of an electromagnetic wave in a random inhomogeneous medium is accompanied by a number of fluctuation phenomena including polarization, fluctuation of phase and fluctuation of amplitude. In the present paper, we concentrate on fluctuation of amplitude.

In a flat universe, because the gravitational potential is time-independent, we can derive a solution of the Maxwell equations (7) and (8) representing monochromatic waves with fixed frequency $\omega$. In other words, we can assume that the electric and magnetic fields have the form $Re(\mathbf{E}e^{-i\omega\tau})$ and $Re(\mathbf{H}e^{-i\omega\tau})$ respectively. In the case of $\lambda = 2\pi\omega^{-1} \ll l_0$, ($l_0$ is the typical length-scale of inhomogeneity for matter in the universe), keeping only first order terms, the wave equation can be simplified as (Feng and Liu 1992)

$$(\nabla^2 + \omega^2)E = -4\omega^2 \phi E. \qquad (9)$$

To solve the equation of wave scattering, in the case of weak fluctuations, we apply the Born approximation to expand E in the series

$$E = E_0 + E_1 + E_2 + \cdots \qquad (10)$$

with $E_0 \gg E_1 \gg E_2 \gg \cdots$. Putting Eq.(10) into Eq.(9), we get two first-order equations:

$$(\nabla^2 + \omega^2)E_0 = 0, \qquad (11)$$

$$(\nabla^2 + \omega^2)E_1 = -4\omega^2 \phi E_0. \qquad (12)$$

If the CBR is perfectly uniformly distributed, there is no net gravitational lensing effect on anisotropies of the CBR. This is a well-known result in geometrical optics and will be confirmed below from the point of view of wave scattering. In order to study the gravitational lensing effect on anisotropies of the CBR, we focus on how the fluctuation part of the CBR is affected. To do so, we separate $E_0$ into a homogeneous part $E_{0h}$ and a fluctuation part $E_{0f}$

$$E_0 = E_{0h} + E_{0f} \tag{13}$$

with

$$E_{0h} = <E_{0h}>; \qquad E_{0f} = E_0 - <E_{0f}> . \tag{14}$$

Here, the averaging is done over all observation directions. As $E_{0h}$ and $E_{0f}$ propagate in an inhomogeneous universe, they interact with the gravitational potential and produce scattered waves $E_{1h}$ and $E_{1f}$ respectively. $E_{0h}$ and $E_{0f}$ and their scattering terms satisfy following equations

$$(\nabla^2 + \omega^2)E_{0h} = 0, \tag{15}$$

$$(\nabla^2 + \omega^2)E_{1h} = -4\omega^2 \phi E_{0h}, \tag{16}$$

$$(\nabla^2 + \omega^2)E_{0f} = 0, \tag{17}$$

$$(\nabla^2 + \omega^2)E_{1f} = -4\omega^2 \phi E_{0f}. \tag{18}$$

After scattering, the outgoing wave is

$$E = E_0 + E_1 = E_{0h} + E_{0f} + E_{1h} + E_{1f}. \tag{19}$$

Because the last scattering surface is far away from us, without loss of generality, we let the incident waves have the plane wave forms, $E_{0h} = A_{0h}e^{i\omega \cdot \mathbf{x}}$ and $E_{0f} = A_{0f}e^{i\omega \cdot \mathbf{x}}$. The scattered waves can then be easily expressed, using Green's function method, in the following forms

$$E_{1h}(\hat{\mathbf{x}}_1) = \frac{\omega^2}{\pi} \int A_{0h}(\hat{\mathbf{x}})\phi(\mathbf{x})\frac{e^{i(\omega x - \omega \cdot \mathbf{x})}}{x}d^3\mathbf{x}, \tag{20}$$

$$E_{1f}(\hat{\mathbf{x}}_1) = \frac{\omega^2}{\pi} \int A_{0f}(\hat{\mathbf{x}})\phi(\mathbf{x})\frac{e^{i(\omega x - \boldsymbol{\omega}\cdot\mathbf{x})}}{x}d^3\mathbf{x}. \tag{21}$$

When Eqs.(20) and (21) are evaluated, it is sufficient to include only the contribution due to waves scattered through angles not exceeding $\theta = \frac{\lambda}{l_0} \ll 1$. In other words, the integration can be confined to the part of space which lies within the cone $C(d\Omega)$, $\theta \leq \frac{\lambda}{l_0}$, where $\theta$ is the angle between the direction of observation and the direction of the scattering element, and $l_0$ is the typical inhomogeneity scale in the universe. In fact, the integration functions of Eqs.(20) and (21) oscillate rapidly outside the cone $C(d\Omega)$, so that for a sufficiently smooth variation of $\phi(\mathbf{x})$, integration over the region external to cone $C(d\Omega)$ only provides a negligible contribution. This cone $C(d\Omega)$ is much smaller than the angular scale of CBR inhomogeneities and so we have

$$E_{1h}(\hat{\mathbf{x}}_1) = A_{0h}(\hat{\mathbf{x}}_1)\frac{\omega^2}{\pi}\int \phi(\mathbf{x})\frac{e^{i(\omega x - \boldsymbol{\omega}\cdot\mathbf{x})}}{x}d^3\mathbf{x}, \tag{22}$$

$$E_{1f}(\hat{\mathbf{x}}_1) = A_{0f}(\hat{\mathbf{x}}_1)\frac{\omega^2}{\pi}\int \phi(\mathbf{x})\frac{e^{i(\omega x - \boldsymbol{\omega}\cdot\mathbf{x})}}{x}d^3\mathbf{x}, \tag{23}$$

$A_{0h}(\hat{\mathbf{x}}_1)$ is the amplitude of the uniform part of the CBR and $A_{0f}(\hat{\mathbf{x}}_1)$ is that of the fluctuation part; as a direct consequence, $E_{1h}(\hat{\mathbf{x}}_1)$ is independent of observation direction, but $E_{1f}(\hat{\mathbf{x}}_1)$ does depend on the observation direction.

### III. ANGULAR CORRELATION FUNCTION

In the Rayleigh-Jeans part of blockbody radiation spectrum, $T \propto I$ (where T and I are temperature and intensity of the CBR respectively) and $I \propto A^2$ (where A is the amplitude of the electric vector E). We will not distinguish A and E in the following except when necessary. We then have

$$\frac{\delta T}{T} = 2\frac{\delta E}{E} = 2\frac{E - \langle E \rangle}{E} \tag{24}$$

Putting Eq.(19) into Eq.(24) and using the definitions of $E_{0h}$ and $E_{0f}$, we have

$$\frac{\delta E}{E} = \frac{E_{0f} + E_{1h} - \langle E_{1h} \rangle + E_{1f} - \langle E_{1f} \rangle}{E_{0h}}. \tag{25}$$

It is obvious from Eqs.(22) and (23) that $E_{1h} - \langle E_{1h} \rangle = 0$ and $\langle E_{1f} \rangle = 0$. Without loss of generality, we let $E_{0h} = 1$ and then the above equation simplifies to

$$\frac{\delta E}{E} = E_{0f} + E_{1f}. \tag{26}$$

Considering the above relations, the angular correlation function of anisotropy of temperature of the CBR may be obtained

$$C(\alpha) = \left\langle \frac{\delta T(\hat{\mathbf{x}}_1)}{T} \cdot \frac{\delta T(\hat{\mathbf{x}}_2)}{T} \right\rangle, \tag{27}$$

where $cos(\alpha) = \hat{\mathbf{x}}_1 \cdot \hat{\mathbf{x}}_2$ and the averaging is done over all observation directions. Substituting Eq.(26) into Eq.(27), it follows that

$$C(\alpha) = \langle |2E_{0f}(\hat{\mathbf{x}}_1)|^2 \rangle + 2\langle |4E_{0f}(\hat{\mathbf{x}}_1)E_{1f}(\hat{\mathbf{x}}_1)| \rangle + \langle |2E_{1f}(\hat{\mathbf{x}}_1)|^2 \rangle. \tag{28}$$

In the right-hand side of Eq.(28), the first term is the angular correlation function of the primordial fluctuation background; the second describes the interaction between the gravitational lensing effect and the perturbed part of the primordial background (this is called the angular cross correlation function $C_c(\alpha)$, and determines the lensing effects on the anisotropy of the primordial CBR); the last is a higher order term and may reasonably be omitted. The major purpose of this paper is to determine the angular cross correlation function $C_c(\alpha)$.

We Fourier decompose the gravitational potential, obtaining

$$\phi(\mathbf{x}) = \frac{1}{(3\pi)^3} \int \Delta_\mathbf{k} e^{i\mathbf{k} \cdot \mathbf{x}} d^3\mathbf{k}. \tag{29}$$

Based on linear perturbation theory and assuming a Gaussian random fluctuation field, we have

$$\Delta_\mathbf{k} = \frac{3}{2} H_0^2 k^{-2} \delta_\mathbf{k}, \tag{30}$$

$$\langle \delta_{\mathbf{k}_1} \delta_{\mathbf{k}_2} \rangle = |\delta_{k_1}|^2 \delta^3(\mathbf{k}_1 - \mathbf{k}_2), \tag{31}$$

$$\langle \Delta_{\mathbf{k_1}} \Delta_{\mathbf{k_2}} \rangle = \frac{9}{4} H_0^4 k_1^{-4} \mid \delta_{k_1} \mid^2 \delta(\mathbf{k_1} - \mathbf{k_2}), \tag{32}$$

$$\langle \delta_{\mathbf{k_1}} \Delta_{\mathbf{k_2}} \rangle = \frac{3}{2} H_0^2 k_1^{-2} \mid \delta_{k_1} \mid^2 \delta(\mathbf{k_1} - \mathbf{k_2}), \tag{33}$$

where $\mid \delta_k \mid^2$ and $\mid \Delta_k \mid^2$ are the spectra of perturbations of the density and gravitational potential respectively.

Roughly speaking, for adiabatic perturbations, the temperature fluctuation of the CBR on scales of a few arcminutes is $\delta T/T = (1/3)\delta \rho/\rho$ before the recombination era (Silk 1967). What are the amplitude and shape of the temperature fluctuations after decoupling? Of course, this depends very much on the assumed recombination history. For the sake of simplicity, we assume that the time-scale of recombination is extremely short so that $\delta T/T = (1/3)\delta \rho/\rho$ remains with the previous amplitude and shape. This assumption is reasonable for our purposes since we are dealing with wave propagation and its interaction with gravitational lensing between the last scattering surface and the observer. Our assumption, in fact, just means choosing a convenient initial condition.

Fourier decomposing the fluctuation part of the CBR and its scattering term, we obtain

$$E_{0f}(\hat{\mathbf{x}}_1) = \frac{1}{6(2\pi)^2} \int \delta_{\mathbf{k}} e^{i\mathbf{k} \cdot \mathbf{x}_1} d^3 \mathbf{k}, \tag{34}$$

$$2E_{1f}(\hat{\mathbf{x}}_1) = \frac{2\omega^2}{3(2\pi)^7} \int \int \int \delta_{\mathbf{k_1}} \Delta_{\mathbf{k_2}} e^{i\mathbf{k_1} \cdot \mathbf{x}_1} e^{i\mathbf{k_2} \cdot \mathbf{x}} \cos(\omega x - \boldsymbol{\omega} \cdot \mathbf{x}) d^3 \mathbf{x} d^3 \mathbf{k_1} d^3 \mathbf{k_2}, \tag{35}$$

where the vector $\mathbf{x}_1$ points to the last scattering surface and has length $2H_0^{-1}$; the unit vector $\hat{\mathbf{x}}_1$ represents the direction of observation.

For mathematical convenience, it is suitable to use a spherical harmonic analysis, which is widely used when dealing with anisotropies on large angular scales:

$$2E_{0f}(\hat{\mathbf{x}}_1) = \sum_{l=0}^{\infty} \sum_{m=-l}^{m=+l} A_{lm} Y_{lm}(\Omega), \tag{36}$$

$$2E_{1f}(\hat{\mathbf{x}}_1) = \sum_{l=0}^{\infty} \sum_{m=-l}^{m=+l} a_{lm} Y_{lm}(\Omega), \tag{37}$$

and

$$A_{lm} = \int 2E_{0f}(\hat{\mathbf{x}}_1)Y^\star_{lm}(\Omega)d\Omega, \tag{38}$$

$$a_{lm} = \int 2E_{1f}(\hat{\mathbf{x}}_1)Y^\star_{lm}(\Omega)d\Omega. \tag{39}$$

Substituting Eqs.(34) and (35) into Eqs.(38) and (39) respectively and using the Rayleigh equation

$$e^{i\mathbf{k}\cdot\mathbf{x}} = e^{ikx\cos\gamma} = \sum_{l=0}^{l=\infty} i^l(2l+1)j_l(kx)P_l(\cos\gamma) \tag{40}$$

and the addition expression

$$P_l(cos\gamma) = \frac{4\pi}{2l+1}\sum_{m=-l}^{m=+l} Y_{lm}(\Omega_{\mathbf{k}})Y^\star_{lm}(\Omega), \tag{41}$$

where $j_l$ and $P_l$ are the *l-th* spherical Bessel function and the *l-th* Legendre function respectively, we obtain the following equations:

$$A_{lm} = \frac{i^l}{6\pi^2}\int \delta_{\mathbf{k}} j_l(2kH_0^{-1})Y_{lm}(\Omega_{\mathbf{k}})d^3\mathbf{k}, \tag{42}$$

$$a_{lm} = \frac{i^{2l}\omega^2}{12\pi^5}\int\int\int \delta_{\mathbf{k_1}}\Delta_{\mathbf{k_2}} j_l(2k_1H_0^{-1})j_l(k_2x)Y_{lm}(\Omega_{\mathbf{k_1}})Y_{lm}(\Omega_{\mathbf{k_2}})$$
$$\frac{\cos(\omega x - \omega\cdot\mathbf{x})}{x}d^3\mathbf{x}d^3\mathbf{k_1}d^3\mathbf{k_2}. \tag{43}$$

According to random field theory, for a Gaussian random field, we have

$$\langle \delta_{\mathbf{k_1}}\delta_{\mathbf{k_2}}\delta_{\mathbf{k_3}}\delta_{\mathbf{k_4}}\rangle = \mid\delta_{k_1}\mid^2\mid\delta_{k_3}\mid^2 \delta(\mathbf{k_1}-\mathbf{k_2})\delta(\mathbf{k_3}-\mathbf{k_4}) + \mid\delta_{k_1}\mid^2\mid\delta_{k_4}\mid^2 \delta(\mathbf{k_1}-\mathbf{k_3})\delta(\mathbf{k_2}-\mathbf{k_4})$$
$$+ \mid\delta_{k_1}\mid^2\mid\delta_{k_2}\mid^2 \delta(\mathbf{k_1}-\mathbf{k_4})\delta(\mathbf{k_3}-\mathbf{k_2}). \tag{44}$$

Combining the above expansions and integrating for angular coordinates within the cone $C(d\Omega)$, we finally obtain

$$<\mid a^2_{lm}\mid> \; = \; <\mid a_{lm}A^\star_{lm}\mid>$$
$$= \frac{4H_0^4}{3\pi l_0^4}\{[\int_0^{2H_0^{-1}}\int_0^\infty \mid\delta_k\mid^2 j_l(2kH_0^{-1})j_l(kx)xdxdk]^2$$
$$+ 2\int_0^{2H_0^{-1}}\int_0^{2H_0^{-1}}\int_0^\infty\int_0^\infty \mid\delta_{k_1}\mid^2\mid\delta_{k_2}\mid^2 j_l^2(2k_1H_0^{-1})j_l(k_2x_1)j_l(k_2x_2)x_1x_2dx_1dx_2dk_1dk_2\}, \tag{45}$$

where the averaging is done over the entire sky and over all observation positions in the universe. Finally, the angular cross correlation function, generated by interaction between the gravitational lensing effect and the primordial perturbed part of the CBR, is as follows:

$$C_c(\alpha) = \frac{1}{4\pi}\sum_{l=2}^{\infty}(2l+1)2<|a_{lm}|^2>P_l(\alpha).  \qquad (46)$$

For a double beam switch experiment, the observable $\delta T(\alpha)/T$, produced by gravitational lensing, is

$$\delta T(\alpha)/T = \frac{1}{4\pi}\sum_{l=2}^{\infty}(2l+1)2<|a_{lm}|^2>_{12}(1-P_l(\alpha)).  \qquad (47)$$

## IV. NUMERICAL RESULTS AND CONCLUSIONS

We assume that the fluctuation spectrum is the Harrison-Zel'dovich spectrum when calculating the gravitational lensing effect on anisotropies of the CBR. For comparison with observation, an appropriate normalization is necessary. We use the rms mass fluctuation $(\delta M/M)^2 = 1$ within a sphere of radius $r_0 = 8h^{-1}Mpc$ as the normalization condition. Of course, how to select normalization conditions depends somewhat on how the mass distribution traces the galaxy distribution. At present, this is not clear and so the selection of the normalization condition may produce an uncertainty in the numerical results. However, this does not significantly affect our conclusion. With the above normalization condition, the normalized power spectrum is given by

$$|\delta_k|^2 = r_0^4 k  \qquad (48)$$

In order to cancel the unknown nonlinear effect, we need to apply a low-pass filter function to truncate the spectrum. For mathematical convenience, we choose this as an exponential form $e^{-kr_t}$. Thus, for making calculations, the power spectrum is replaced by

$$|\delta_k|^2 = r_0^4 k e^{-kr_t}.  \qquad (49)$$

Here $r_t$ is a cut-off scale. Integrating Eq.(45) over k, we obtain

$$\langle a_{lm}^2 \rangle = \frac{2(H_0 r_0)^4}{3\pi} (\frac{r_0}{l_0})^4 \{[\frac{H_0}{4} \int_0^{2H_0^{-1}} Q_l(\frac{r_t^2 + 4H_0^{-2} + x^2}{4H_0^{-1} x}) dx]^2 +$$

$$\frac{H_0^2}{8} Q_l(\frac{1 + r_t^2 H_0^2}{8}) \int_0^{2H_0^{-1}} \int_0^{2H_0^{-1}} Q_l(\frac{r_t^2 + x_1^2 + x_2^2}{2 x_1 x_2}) dx_1 dx_2 \} \quad (50)$$

with

$$l_0 = \frac{\int \frac{1}{k} |\delta_k|^2 d^3\mathbf{k}}{\int |\delta_k|^2 d^3\mathbf{k}} = 3 r_t, \quad (51)$$

where $Q_l$ is the second kind of Legendre function. We have integrated Eq.(50) numerically in the case of $r_t = 5 h^{-1} Mpc$, $r_0 = 8 h^{-1} Mpc$ and $h = 0.75$. The behavior of the multipoles as a function of the harmonic number l is showed in Fig.1. The predicted cross function of the temperature fluctuation of the CBR due to the gravitational lensing effect for comparison with double beam switch experiments is showed in Fig.2. We have also made numerical calculations varying the parameters $r_t$, $r_0$ and h within acceptable regions. However, the numerical results are not sensitive to changes of these parameters. From Fig.2, it is obvious that the effect of gravitatioal lensing on anisotropies of temperature of the CBR on the scale of arcminutes are too small to significantly amplify or depress the primordial anisotropies of the CBR. We can then safely conclude that the gravitational lensing effect on anisotropies of temperature of the CBR on scales of a few arcminutes are negligible. Our conclusion disagrees strongly with that of Cayon et al. who argue that gravitational lensing, produced by linear density perturbations, may enhances $\Delta T/T$ by $\sim 20\%$ or even more.

In addition, the problem in the calculation presented by Cayón *et al.* (1993) could be that they considered the effect on the anisotropies of the CBR caused only by deflections of light lines, but not by convergences and divergences of light beams. In principle, the last effect could also produce a considerable change in the fluctuation of the intensity of the CBR compared with that caused by deflections of light lines.

Acknowledgements.

We thank the anonymous referee for his enlightening comments which resulted in important improvements of the paper. We appreciate Prof. J.Miller for his carefully reading our manuscript which improved its readability. J.M.L. is grateful to Dr. L.Toffolatti, Dr. P.Mazzei, Dr. N.Di.Cicco and especially to Prof. G.De Zotti and Prof. L.Danese for their hospitality during his stay in Padova. He also acknowledges financial support from SISSA and from the Osservatorio Astronomico di Padova where part of this work was done.

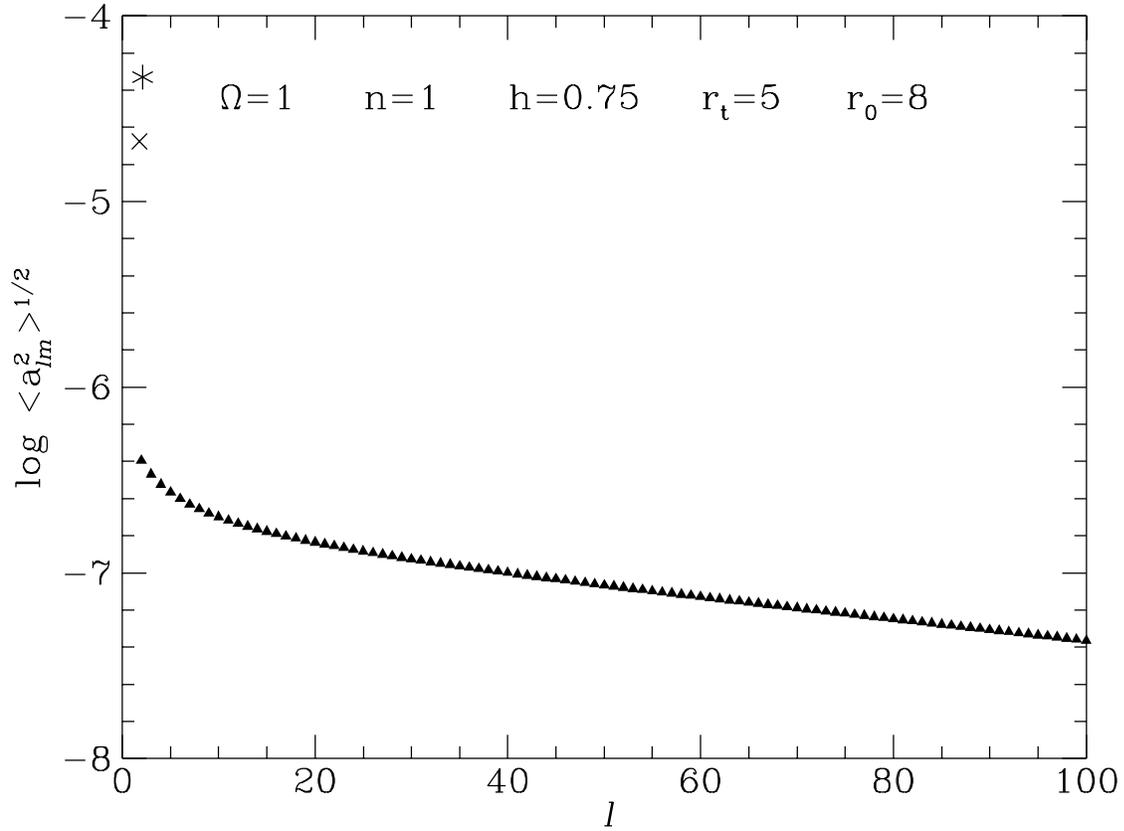

Fig. 1: The hamonic amplitude of the secondary fluctuations produced by the gravitational lensing as a function of $l$. $\times$ and $*$ represent the amplitude of the quadrupole moment of primary fluctuations for h=1.0 and h=0.5 respectively (Holtzman 1989).

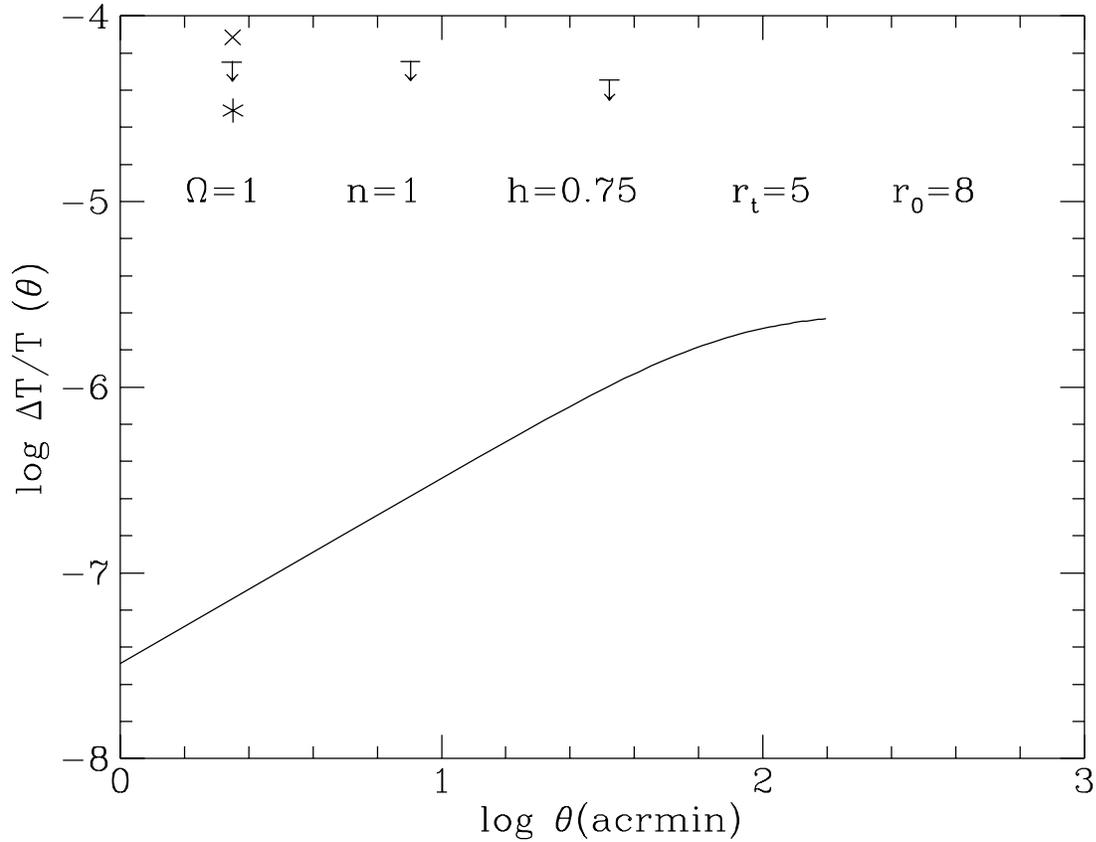

Fig. 2: The secondary temperature fluctuations as a function of angular. × and ∗ represent the primary temperature fluctuations at 4.5' for h=1.0 and h=0.5 respectively (Holtzman 1989). The arrows at 4.5', 7.2' and 30' show the upper limits of Uson & Wilkinson (1984), Readhead et al. (1989) and Gaier et al. (1992).